# Electrocorticographic Dynamics Predict Visually Guided Motor Imagery of Grasp Shaping

Jing Wu, *Student Member, IEEE*, Kaitlyn Casimo, David J. Caldwell, *Student Member, IEEE*, Rajesh P.N. Rao, *Member, IEEE*, and Jeffrey G. Ojemann

*Abstract*— Identification of intended movement type and movement phase of hand grasp shaping are critical features for the control of volitional neuroprosthetics. We demonstrate that neural dynamics during visually-guided imagined grasp shaping can encode intended movement. We apply Procrustes analysis and LASSO regression to achieve 72% accuracy (chance = 25%) in distinguishing between visually-guided imagined grasp trajectories. Further, we can predict the stage of grasp shaping in the form of elapsed time from start of trial ($R^2=0.4$). Our approach contributes to more accurate single-trial decoding of higher-level movement goals and the phase of grasping movements in individuals not trained with brain-computer interfaces. We also find that the overall time-varying trajectory structure of imagined movements tend to be consistent within individuals, and that transient trajectory deviations within trials return to the task-dependent trajectory mean. These overall findings may contribute to the further understanding of the cortical dynamics of human motor imagery.

## I. INTRODUCTION

The ability to translate the cortical activity of imagined motor movements into usable motor output is a crucial step in the development of volitional neuroprosthetics [1]–[3]. Electrocorticography-based brain-computer interfaces have previously demonstrated promise in reconstructing overt reach-to-grasp movements [4]–[9], as well as to control brain-computer interface tasks both with and without overt movement [10]–[14]. However, these previous decoding approaches are limited by their attempts to find the neural direct correlates of overt movement performed in motor-intact individuals, or by their reliance upon an extensive BCI training process through which subjects gradually learn to modulate task-specific cortical activity with specific abstract feedback such as center-out cursor movements. While cortical spectral differences from rest related to motor imagery can be observed in the sensorimotor cortex of untrained BCI-naïve individuals, they are reported to be substantially weaker in magnitude [14], and subsequently less successfully suited as features for real-time movement decoding, prior to protracted BCI feedback training.

However, while direct spectral power changes associated with movement imagery may be small in magnitude in the primary motor cortex, the overall time-varying dynamics of the cortical signal from both motor and non-motor regions may still bear valuable information with regards to motor planning and intent. Recent investigations in neural populational dynamics during reaching movements reveals that even noisy single-trial dynamics of neural populations can provide accurate state estimations given the correct dynamical model [15], [16]. These findings signify that time-varying neural trajectories, when projected into a subspace that maximizes output-relevant dimensions, can extract task-specific movements even when mixed within signals arising from task-irrelevant processes or noise.

A dynamical model incorporating broad anatomical regions to reconstruct motor imagery is also particularly relevant in light of previous work establishing that parietal and premotor cortices play a key role in the representation of reach movements [17]–[19]. Patients with ECoG electrodes implanted for medically intractable epilepsy provide recordings over a large brain region, providing the spatial coverage that may be necessary for full reconstruction of the underlying motor imagery dynamics, which can overcome some generalization limitations in prior studies due to limited anatomic scope.

In this work, we investigated ECoG recordings in three patients with implanted electrode grids. The patients performed overt and imagined motor movements synchronous to visual observation of an animated hand. Using a dynamical model based on Procrustes analysis, we examined the movement type and timing of imaginary movements for the prediction of grasp shaping and timing, achieving high classification accuracy with only dozens of trials in BCI-naïve individuals.

## II. METHODS

### A. Experimental Task and Data Acquisition

Three patients (ages: 13-33; 1 female, 2 male; 2 left hemisphere, 1 right) with normal motor function volunteered for this study during clinically recommended invasive monitoring of seizure-related cortical regions, with written, informed consent through a protocol approved by the Institu-

Research supported by grants from NSF EEC-1028725, NINDS 5R01NS065186, and the WRF Fund for Innovation in Neuroengineering.

J. Wu is with the department of Bioengineering, University of Washington, Seattle, WA 98105, USA, and the NSF Center for Sensorimotor Neural Engineering (CSNE). (phone: +1-206-685-3301; e-mail: jiwu@uw.edu).

K. Casimo is with the Graduate Program in Neuroscience, University of Washington, Seattle, WA 98105, USA, and the CSNE. (email: kcasimo@uw.edu)

D.J. Caldwell is with the department of Bioengineering, University of Washington, Seattle, WA 98105, USA, and the CSNE. (email: djcald@uw.edu)

R.P.N. Rao is with the department of Computer Science, University of Washington, Seattle, WA 98105, USA, and the CSNE. (e-mail: rao@cs.washington.edu).

J.G. Ojemann is with the department of Neurosurgery, University of Washington, Seattle, WA 98105, USA, and the CSNE. (e-mail: jojemann@uw.edu).

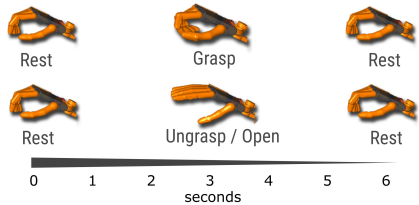

Figure 1. Two of four example grasp shaping trial visual stimuli shown: Grasp and Ungrasp. Not shown: Pinch, Unpinch.

tional Review Board. All subjects subject received a pre-operative anatomical MRI, and was implanted with an 8×8 platinum electrode grid (Ad-Tech, Racine, WA, USA), along with 1×8 and 1×6 electrode strips (excluded from this analysis). All electrode grids and strips consist of 3mm diameter platinum discs, set in silastic with center-to-center distances of 10mm. Subsequently, X-ray CT imaging captured the location of the electrodes. These electrodes were sampled at 1220.7 Hz on a 128-channel Tucker-Davis Technology (Alachua, FL, USA) System 3 RZ5D Neurophysiology Base Processor with associated PZ5 NeuroDigitizer.

The patients were comfortably positioned in front of one screen during the task. The screen displayed two separate videos of an animated robotic hand [20] with four different smooth movements, consisting of two pairs of kinematically opposing movements: a power-grasping movement (*grasp*), a spherical-open movement (*ungrasp*), a palmar pinch movement (*pinch*), and an claw-open movement (*unpinch*). One of these movement pairs (*grasp-ungrasp*) are shown in Fig 1.

Each of these smooth movements began with a neutral pose (0s), proceeded to the designated pose (3s), and returned to a neutral pose (3s). The patients carried out the poses using the hand contralateral to their implant, regardless of hand dominance. Patients were visually cued to alternately follow with their own hand, or with motor imagery only. In the imagined movement condition, patients were instructed also to imagine feeling the sensations of the movement animation as they would with overt movement. Each animation proceeded for 6s per trial with a 2s inter-trial intervals. Each patient performed 20 to 70 movement trials.

### B. Data Preprocessing and Feature Extraction

The pre-operative anatomical MRI was co-registered with the postoperative ECoG CT using Statistical Parametric Mapping (www.fil.ion.ucl.ac.uk/spm), with cortical surfaces reconstructed with FreeSurfer (freesurfer.net) and custom mapping and projection code implemented in MATLAB (Natick, MA, USA) [21]. The recorded channels from the implanted grids were filtered with a common average reference filter within each implanted grid. The filtered signals were then analyzed for time-frequency content by continuous wavelet transform with the non-analytic Morlet wavelet defined by $\Psi(s\omega) = \pi^{-1/4} e^{[-(s\omega-\omega_0)^2]/2}$, where $\omega_0=6$ and $s$ reflects log scaling factors to obtain ¼-octave resolution across pseudo-frequencies 2−200 Hz. We used the absolute magnitude of the wavelet coefficient time series at each scaling. All time series were binned by 50ms and normalized to unit variance per captured channel-frequency.

### C. Procrustes Analysis Classification

The filtered and normalized wavelet coefficients, constituting a time series for each channel × frequency pair, was dimensionally reduced using principal component analysis (PCA) to the top 20 principal components (PCs). The time series were then projected into this 20-dimensional space.

Procrustes analysis was employed for classification by statistical shape similarity. The shape of the dynamical signal trajectories of each of 8 grasp conditions: overt- (grasp, ungrasp, pinch, and unpinch) and imagined (grasp, ungrasp, pinch, and unpinch) were analyzed, and for each individual recording session, leave-one-out validation was applied. For all training trials, the mean trajectory templates were computed for each grasp condition. These mean time series were used as templates against which per-condition Procrustes distance was computed for each testing trial. Fig. 2 illustrates the category-mean template trajectories for each individual. The condition with the minimum Procrustes distance bears the maximum 20-D shape similarity to the testing trial, and is classified as that grasp condition.

### D. LASSO Movement Phase Classification

For all visually-guided motor imagery trials we further employed least absolute shrinkage and selection operator (LASSO) regression [22] to determine the phase of the imaginary smooth grasp shaping movement. The 20D mean trajectory of each imagined grasp condition computed from training trials was used to fit linear regression coefficients to predict seconds of elapsed time. The Lasso regularization parameter was chosen such that the mean squared error is lowest in 6-fold cross-validation. Testing trials were multiplied by the regression coefficients to obtain elapsed time since the start of trial to obtain predicted movement phase.

### III. RESULTS

#### A. Grasp Shape Classification Using Procrustes Analysis

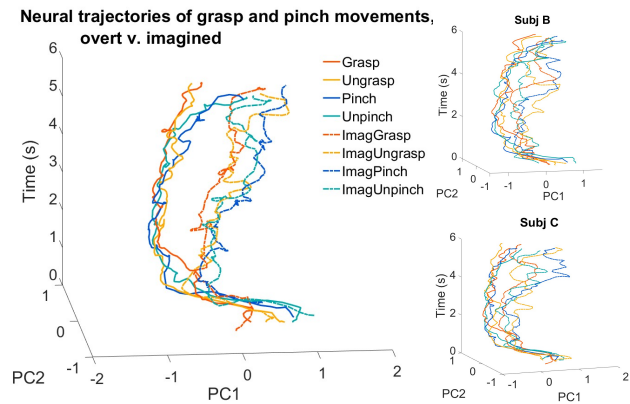

Figure 2. Plot of the mean time-varying cortical power spectrum varying through time. Top two PCA components shown in x-y plane. Left: Subject A. Right Top: Subject B. Right bottom: Subject C.

![Confusion matrix]

Figure 3. Confusion matrix of the model classification accuracies using Procrustes anlaysis using leave-one-out validation, in 136 trials in 3 subjects. Left: actual conditions. Bottom: Predicted conditions. Main diagonal: correctly predicted trials. Off-diagonal: incorrect predictions.

Dynamical trajectory analyses reveal separation between grasp types in dimensionally reduced wavelet transform ECoG recordings, as well as trajectory similarities in closely-related grasp postures (see Fig. 2). In this case, imaginary trajectories were classified at a higher success rate than overt movement trajectories under Procrustes analysis. Mean true-positive classification accuracy was 68% for all conditions (12.5% chance), and mean imaginary grasp shaping classification accuracy was 72% between 4 conditions. The results of the classification are represented in the confusion matrix in Fig. 3.

### B. Movement Phase Regression and Prediction

Cross-validated movement elapsed time agree with predicted using Lasso regression indices in all subjects, with mean $R^2 = 0.4$ across the 3 subjects. This reflects the degree to which each moment within the visually-guided imagined movement is unique in dimensionally-reduced space, and phase can be identified in time by regression alone. Fig. 4 illustrates an example of a mapping between the trajectory and elapsed time, and predicted elapsed trial time.

### IV. DISCUSSION

#### A. Statistical Shape Similarity as Stable Dynamical Feature

In this work, we used Procrustes distance from the mean template trajectory as a proxy measure to classify between dimensionally-reduced, frequency-normalized wavelet coefficients. We demonstrate high classification performance in distinguishing four different types of both overt and imagined single-trial smooth grasp shaping movements. The success of this relatively simple statistical shape metric carries implications about the nature of the underlying signal dynamics under visually-guided motor imagery:

First, this signifies that the overall trajectory shape of the cortical signal remains relatively stable, despite the existence of relatively large trial-to-trial variances in specific timespans of the trajectory. While each individual trial may exhibit deviations from the expected mean template in certain time segments within a trial, the deviations tend to return to the template trajectory, such that the overall shape of the multidimensional trajectory still more closely approximates its expected overt or imaginary grasp condition, allowing for ordinary Procrustes distance to correctly identify the class.

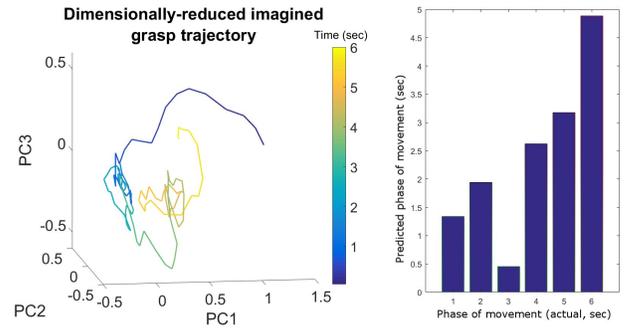

Figure 4. Left: Sample mean trajectory for imagined-pinch for Subj A projected into the top 3 PCs. Color reflects time elapsed in trial. Note relatively unique mapping for each timepoint in 3-PC space. Right: Sample binned predicted timestamps for the nearest second in a single trial.

Second, this result may further imply that task-relevant cortical activity exists in a high-dimensional subspace that is captured by the overall shape of the time series formed by principal component projections of frequency-normalized wavelet coefficients, and that this task-relevant manifold can be relatively resistant to non-task-related spectral changes. While the combination of PCA and Procrustes analysis are not specifically designed to separate task-relevant and task-irrelevant manifolds of cortical activity, the time-locked structure of this visually-guided task may force cortical trajectories perturbed by task-irrelevant activity to return to the nearest task-relevant state, resulting in a preservation of trajectory structure.

Further work into the structure of these dynamics should incorporate techniques that are specifically suited for separating latent factors, such as Gaussian-process factor analysis [23]. Furthermore, specific insights about the task-relevant stability of these trajectories should consider rotation, scaling, and shifting as different manipulations to the dynamical structure of the cortical signals, and their significance with respect to the neural state. For example, trajectory rotations would imply a change to the manifold of underlying cortical signal modulation.

#### B. Somatotopic Localization of Principal Components

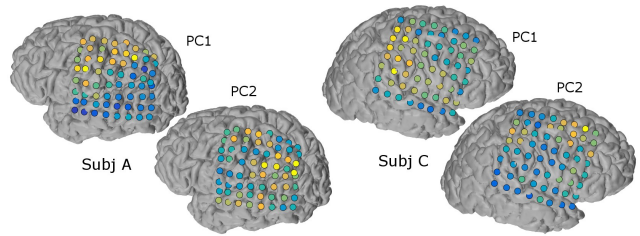

Figure 5. Example localization for the top PC weights for two subjects. Contributions from all frequency bins have been summed (root sum of squares) for spatial visualization. Blue: low contribution to PC. Yellow: high contribution to PC. Note this 2D representation collapses frequency and cannot represent the full range of trajectory features subsequently used in Procrustes analysis.

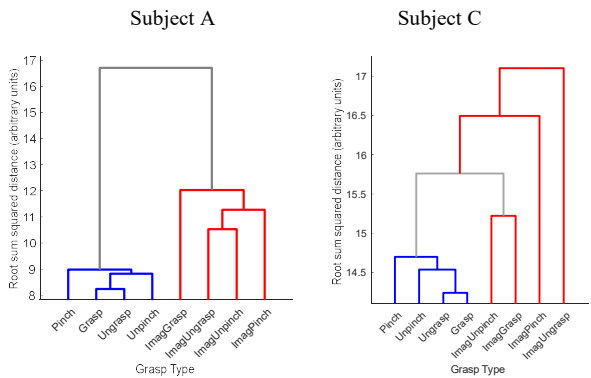

Figure 6. Grasp trajectory similarity dendrogram computed as the sum squared euclidian distances across time between mean grasp condition template trajectories for left: Subj A, right: Subj C. Even with varying electrode coverage, different individuals have very similar overt grasp similiarity hierarchies, but very different imaginary similarity hierarchies.

The top PCs of all recorded electrodes in all subjects involve sensorimotor cortex, premotor cortex, and posterior parietal regions when electrode coverage is available, as expected. However, other PCs involve regions not traditionally associated with motor movement, such as dorsal temporal regions or the temporoparietal junction. These may reflect task-dependent modulation in visual attention during the task, or other unidentified internal representations of motor imagery warranting further investigation.

Further work may involve identifying transient functional connectivity changes between these regions, and identifying similar clusters of spatial-spectral interactions between regions during guided visual imagery.

*C. Movement Similarity Hierarchies*

As can be both observed in Fig. 2 and Fig. 6, there is a strong similarity within the overall trajectories of all overt movements, but very different dendrograms are seen between visually-guided imaginary movements. Some of this variability can be attributed to different electrode placement, but the retained high classification rate implies that despite varied placement, different individuals retain structural consistency in individual time-varying representations of imaginary grasp-shaping behavior, in a wide variety of motor and non-motor regions.

ACKNOWLEDGMENT

We would like to acknowledge Luke Bashford who provided valuable feedback during early analyses of our task.